\documentclass[12pt]{article}
\begin{document}
\def\thebibliography#1{\section*{REFERENCES\markboth
 {REFERENCES}{REFERENCES}}\list
 {[\arabic{enumi}]}{\settowidth\labelwidth{[#1]}\leftmargin\labelwidth
 \advance\leftmargin\labelsep
 \usecounter{enumi}}
 \def\newblock{\hskip .11em plus .33em minus -.07em}
 \sloppy
 \sfcode`\.=1000\relax}
\let\endthebibliography=\endlist

\hoffset = -1truecm
\voffset = -2truecm

\title{\bf Foundations Of Quantum Theory Revisited \footnote {To be published as part of a book " India in the World of Physics : Then \& Now " , A.N. Mitra ed, Pearson Education New Delhi, 2007-08} }
\author{
{\normalsize 
\bf A.N.Mitra \thanks{e.mail:
ganmitra@nde.vsnl.net.in } 
}\\
\normalsize  244 Tagore Park, Delhi-110009, India }
\date{}
\maketitle



\begin{abstract}

After giving a panoramic view of the " text-book "   interpretation of the new quantum 
mechanics, as a sequel to the old quantum theory, the conceptual basis of quantum  
theory since the Copenhagen Interpretation is reviewed  in the context of various 
proposals  since  Einstein and Niels Bohr,   designed to throw light  on  
possible new facets bearing on its foundations, the key issues to its inherent 
" incompleteness "   being  A) measurement, and  B) quantum non-locality.  A related 
item on measurement, namely Quantum Zeno (as well as anti-Zeno) effect  is 
also reviewed briefly. The inputs  for the new facets  are from  some key Indian experts : \\
S.M. Roy, V. Singh ; D. Home; B. Misra; C.S. Unnikrishnan

\end{abstract} 

\section{Quantum Theory : Standard Picture } 

The  issues concerning the foundations of quantum theory  mainly hinge on two aspects, namely, A) quantum measurement problem, and  
B) quantum non-locality. To appreciate these subtleties, it is  first necessary to  have  an  overview  of  its  "standard  picture"  [1]
which is  summarized below.   
\par    
After the discovery by Max Planck  of the existence of discrete quanta to account for the 
observed black-body spectrum, Quantum Theory got a big boost  at the hands of Einstein 
himself  who found a natural explanation of some crucial experimental observations  on  
the photoelectric effect in terms of the particle nature  of  electromagnetic radiation,  thus 
establishing  its  $dual$ character. What was now needed was a  viable atomic model  to 
account for  the discrete values of several measurable parameters of  atomic systems, 
especially i) the Ritz classification of spectral lines in terms of the Rydberg - Ritz combination  
principle;  and ii) Franck-Hertz experiment on the discrete energy losses of electrons on  
collision with atoms.  To that end, the right background was  provided  by  Rutherford's 
discovery of the atomic structure, and a crucial step was taken by   Niels Bohr  who  
postulated that  i) atomic systems can only exist  in  certain quantized states,  each  
corresponding to a well- defined energy, so that transitions between them are accompanied  
by radiation whose energy ($E$) equals  their  energy difference  ; and ii) the frequency  
of the radiation quantum is equal to  $E / \hbar$.   These two postulates sufficed  for an 
understanding of  both the  Rydburg-Ritz combination principle, and the Franck-Hertz  
experiment.  Bohr's postulates received a further boost  through a quantization rule  
discovered by Wilson and Sommerfeld within the Hamilton-Jacobi formulation of classical 
mechanics,  namely, the (classically derived) action integrals must be integral multiples of 
$\hbar$,  so that  the corresponding energy levels, expressed  in terms of the action integrals,  
should automatically have quantized values. This  so-called  " Old Quantum Theory "  was  
highly successful in explaining  a huge mass of  spectroscopic data , such as   the fine 
structure of the hydrogen atom, the spectra of diatomic molecules,  without further assumptions. 
But there were many difficulties  in the way of a `natural'  understanding of the Old Quantum Theory.  

\subsection{Difficulties with Old Quantum Theory  }

The  difficulties [1]  were both i) practical and ii) conceptual. i) Practical,  because this theory  was not 
applicable to aperiodic systems, not did it  properly account for  the $intensities$ of spectral  
lines, and with improvement in experimental techniques, the gap between experiment  and 
theory  increased. [ The Correspondence Principle was introduced by Bohr to ensure better   
agreement  in the limit of large quantum numbers  when the classical conditions are more 
valid, but it was at most a stop gap arrangement].  
  ii) Conceptual, because a conceptually 
satisfactory explanation of  the basic  phenomena was lacking. [ For example it was difficult to 
understand why the Coulomb  force  in the hydrogen atom was so effective for the spectroscopy,  
while the ability of an accelerated electron to emit radiation disappeared in a stationary state].  
And the assumption of a dual character of light (particle-like on emission and absorption,  and 
wave-like in transit) seemed  to lack logical self-consistency.  Most of these problems 
disappeared with  a more elaborate  approach under the name of   New Quantum Mechanics.  

\subsection{New Quantum Mechanics}

The advent of the New Quantum Mechanics was preceded by several conceptual breakthroughs 
in quick succession.  First,  de Broglie (1924)  showed through his postulate  
$\lambda = h / p $ that  wave - particle  duality  was not a monopoly of radiation alone, as    
emanating  from the  Planck-Einstein discoveries.  This concept was equally applicable  to material 
particles, as was  soon to be  demonstrated by the Davisson-Germer (1927) and  G.P. Thompson  
(1928) experiments. Secondly  S.N. Bose (2)  introduced the concept of $indistinguishability$ 
through a new mode of counting  for the derivation of Planck's law, a result which, as noted above, 
was endorsed by Einstein through a  corresponding  derivation  for material particles [3].  The     
concept  of  indistinguishability  which  stemmed from the Bose form of counting,  had no   
"classical" counterpart, and led to the prediction by Einstein   of Bose-Einstein condensate, one in which a finite fraction 
of particles trickle down to the lowest state [3].  Bose's concept also found a natural echo  in  the Matrix Mechanics  
of Heisenberg who  abandoned the concept of individual particle identity  that   had  
characterized  classical mechanics, in favour of $operators$  (as matrices) for dynamical  
variables like position ($q$) and momentum ($p$) on the one hand,  and $states$ (as vectors)  
on which  the dynamical variables operate to produce their "measured"  values  (in a  
representation labelled by   suitably discretized variables) on the other.  Thus the individual 
values of dynamical variables in classical mechanics now gave way to an $array$ of numbers 
(the matrix elements)  in their matrix representation. In the new approach, the problem of 
finding the $values$ now reduced to one of `diagonalization' of these matrices, the diagonal 
elements being interpreted as the only results of $measurement$,  viz., $eigenvalues$, and 
the corresponding  vectors being interpreted as the  possible $states$, viz., $eigenstates$,  
on which the measurements were valid.  This concept of  $measurement$  in turn necessarily  
led to  the  (dual) concept of  object- observer pair,  thus vastly extending the domain of the  
physics beyond the corresponding classical  domain  which had no scope for  an observer  
as a separate dynamical entity.  

\subsection{Uncertainty Principle and Complementarity}

The outcome of this new  formalism  [1,4-5] may be summarized by the so-called $Uncertainty$ 
$Principle$ ($UP$)  of Heisenberg (1927), according to which it is impossible to specify precisely  
and simultaneously the values of both members of $canonically$ $conjugate$ pairs of  
dynamical variables (like $(q, p)$) that describe the behaviour of atomic systems. Some 
other pairs of canonically conjugate variables, in an obvious notation,  are $(J_z , \phi)$ 
and $(E, t)$.  The uncertainty relations for  typical pairs  in an obvious notation  are :
$$  [ \Delta x \times \Delta p_x ; \Delta \phi \times \Delta J_z; \Delta t \times \Delta E ] \geq \hbar/2 $$ 
which imply that  the determination of any dynamical variable with unlimited precision must 
be traded with a corresponding loss of information of its canonically conjugate momentum, 
and vice versa. And the smallness of $\hbar$ makes the result of physical interest only  
for systems of atomic size or less. In particular, an  energy  determination with accuracy 
$\Delta E$ must cost a time interval of at least $\sim \hbar /2 \Delta E$, which implies, e.g.,  that 
energy conservation in a certain process can be violated with impunity provided the time 
available for measurement is short enough ! As a word of caution, however,  the   time--energy 
UP should be distinguished  from the other UP's, since  a formal 
time operator is not defined in conventional quantum mechanics, as first shown by  Pauli. 
\par 
For a physical understanding of the implications of $UP$, Bohr introduced the $Complementarity$ 
$Principle$ which states that atomic phenomena cannot be described with the completeness 
demanded by classical mechanics. Now certain pairs (the canonically conjugate ones !) that  
complement each other  for a complete classical description, are actually mutually exclusive, 
although they are needed for a full description of various aspects of  the phenomenon.  
On the other hand, from the experimenter's point of view, the complementarity  
principle  asserts that the physical apparatus available to him  is so constrained  that more 
precise measurements than those  mandated by the $UP$ cannot be made.  This must not be regarded  
as  a limitation of the experimenter's techniques, but a more intrinsic law of nature which dictates that 
whenever an attempt is made to measure one of a pair of canonical variables, the other is changed 
by an amount that cannot be  estimated  without interfering with the primary attempt.   
This is fundamentally different from the classical situation, in which the measurement process 
in principle disturbs the system under observation no doubt, but the amount of disturbance is 
either too small to be of consequence, or it can be calculated and taken into account;  in either 
case the `disturbance' due to measurement is $not$  of  much $dynamical$ significance. Not so for 
an atomic system,  whose behaviour cannot be described independently of the means by which 
it is observed, so that the object and the observer are inextricably linked  together  in a 
dynamical  fashion. 
\par
Stated differently, the situation for an atomic system is as follows : One must choose between various  
experimental arrangements, each designed to measure the two members of a pair of canonical 
variables with different degrees of precision that are compatible with the $UP$. In particular, there 
are two extreme arrangements, each of which measures one member of the pair with great 
precision. In  the `classical' theory,  these extreme experimental arrangements  complement 
each other; the results of both  are available simultaneously, and indeed are necessary for 
a complete description of the system.  But in  a `quantum'  theory which  actually applies to the 
atomic  system,   such extreme complementary experiments are mutually exclusive,  and cannot 
be performed simultaneously.  Thus the classical concept of causality  is no longer valid in a 
quantum situation: There is causality to the extent the  quantum equations of motion are 
perfectly well defined. But the causal relationship between successive configurations of an 
atomic system that characterize a classical description, no longer exists. And  the role 
of $measurement$  is now an  active  part of the  quantum dynamics.    
  
\subsection{Simple Diffraction Experiment [1, 5] }

The way New Quantum Mechanics " resolved " the inner contradictions of the Old Quantum Theory 
is  best illustrated by  the  interpretation  of a simple diffraction experiment  which though 
text-book material [1],  has acquired  a  renewed significance [4, 5] in  the   context 
of recent developments on the  very foundations of  modern quantum theory.  
\par
Consider a  double-slit  experiment  of the standard type [1, 5] (figure omitted for brievity, but  
the notation of [5a] is kept in the following).   
A light source S  is placed in front of a screen  A in which two parallel slits $A_1, A_2$  are cut in horizontal 
directions; a second screen M parallel to (and behind ) A, is equipped with proper devices to measure 
the  pattern.  Now consider three different cases: 1) particles; 2) waves; 3) electrons as follows [5]. 
\subsubsection{Particles}
When the source shoots (classical) point particles, say bullets, measure the vertical distribution $P$ 
of  the fraction of the number of bullets arriving at M. In this case each bullet goes through either 
slit $A_1$, or through $A_2$, and arrive at a definite point on the screen M. Let the vertical distribution 
be $P_1$, if $A_1$ is open; $P_2$,  if $A_2$ is open; and $P_{12}$, if both are open. In this case,  
$P_{12} = P_1 + P_2$, a  commonsense result which in the modern language goes by the name of a  " decoherence "  effect  [5b]
 due to multiple interactions of the `bullet'  with the environment, resulting in a clear `which slit ' passage for its  motion (see below).   
\subsubsection{Waves}
Next consider water waves  produced at the source S, and measure the vertical  
intensity $I$ of the wave motion which takes on the values $I_1, I_2, I_{12}$ for the three 
cases as above, respectively. This time a full-fledged (albeit classical)  wave-like scenario ensures a typical 
interference pattern ( via the principle of superposition)  for the resultant intensity distribution  when both slits are open, so that 
$I_{12} \neq I_1 + I_2$.  
\subsubsection{Photons }     
Now consider the source S emitting radiation , to illustrate the quantum scenario.     
 Here we have the radiation  
behaving like a wave during its passage from the source S to the screen B  via the slit A,  but 
behaving  like quanta (photons) when ejecting photoelectrons from B, so that the wave and 
particle aspects appear in the same experiment.  What is the distribution now ? We find a 
diffraction pattern, namely $I_{12} \neq I_1 + I_2$ ! 
 Is it due to the interference between different photons passing through  the 
two slits ?  This explanation is not sufficient, since the diffraction pattern still appears when 
the intensity of light is so much reduced that on average, only one photon at a time passes  
from the source to the screen, which must be through one of the two slits only.  So one must 
conclude that the diffraction pattern is a statistical  property of a $single$ photon, and $not$  
due to interference of more than one photon.   This leads one to  ask: " how does the presence of  
a slit    through which the photon does not go, prevent it from reaching a part of the  
screen that it  would be likely to reach if that slit were closed ?" [1]
\par
Quantum mechanics " resolves "  the issue [1]  through the assertion that the diffraction pattern would 
be destroyed if a sufficiently careful attempt were made to determine through $which$ slit each 
photon passes. Thus if we place a detector C near one of the  slits to find out if it passed through  
that slit, we find indeed that  $P_{12} = P_1 + P_2$, i.e., the diffraction pattern is destroyed  [1] ! 
In the modern language of Decoherence which  may be defined as a process by which the environment  destroys the 
wavelike nature of things by getting information about a quantum system [5b],  the interaction with the detector C introduces a `decoherence' effect  resulting in a clear `which-slit'  passage  for the electron. But in the absence of this decoherence,  the diffraction pattern is restored 
as if the photon  were a wave only.      A more conventional explanation  is that any attempt  to find out through which slit the photon  has passed gets into direct conflict with the Uncertainty Principle [1]. 

\section{ Anatomy Of Quantum Theory : New Facets} 

The foregoing is a text-book  background [1]  on the ramifications of  Standard Quantum Mechanics 
vis-a-vis the classical theory, illustrated with a familiar diffraction experiment.  Before going into  
its  depths, it is  perhaps in order to  stress  that the first  victim of the quantum  paradigm  was the  
Cartesian  Partition between the $physique$ and the $psyche$, since  a $measurement$ (involving  
as it did,  a close  interaction between the two)  now  became a key  ingredient  of the new physics,  
which  severely  restricted    the   hitherto "ontological" (out there) status of a classical observable.    
Initially therefore it was quite hard for the Western physics  community to  adjust  to  this sudden  
change in paradigm. This was best illustrated by  Einstein's  profound unhappiness [6] with the  
`incompleteness' of quantum mechanics as revealed,  for example, by the diffraction experiment.  
 Curiously enough, it  did not seem to sound  such an unfamiliar ring with   "eastern thought" ,  
with   its penchant for mysticism (!), as  was once  to  be revealed  to  L. Rosenfeld during his  
discussions with Japanese physicists on the subject.  Einstein's famous conversations with 
the poet  Rabindranath Tagore also pointed to a similar  divergence of emphasis between 
the ontological and philosophical   aspects of reality  ( for details see Home [4]).   
\par
The other aspect of the foundations of quantum theory is that while such studies were for  long  
regarded  by many to be of "metaphysical" importance only [4], have suddenly sprung up 
over the last 2-3 decades  as frontline items of study, thanks to the growth  of new 
experimental techniques on the one hand, and to the development of {\it quantum technologies 
of information processing and transfering} on the other . This has led to spectacular advances  in quantum optics technology and quantum 
information processing.   Therefore it is of great physical interest to investigate the 
manner in which the abstract conceptual issues of quantum mechanics can be linked to the  
actual experiments in order to obtain new insights, as well as to uncover some of the 
unexplored facets of quantum mechanics.    

\subsection{ Classical Physics Ontology [5]} 

To bring a semblance of order to the turbulence caused by the sudden paradigm shift,  it is 
useful  to  have a critical  reassessment of  the  classical  premises prior to the quantum formulation. 
Two basic hypotheses which the Greek thinkers made  about Nature, according to Schroedinger  
(quoted in Singh [5a])   
are:  1) the existence of a real  external world, which amounts to taking the observer's  
consciousness out of the purview of the  observed world; 2) this external world is accessible 
through the existence of laws of nature (read Newton's laws of classical mechanics).   
\par
Next one must emphasize the $unitary$ nature of classical mechanics, namely  [5a], it describes  
both the system under observation, the measuring apparatus as well as their mutual  interaction. 
The interaction disturbs the system in principle, but the disturbance can be reduced to any desired  
level of accuracy, so that the measurement is not an epistemological problem, rather a practical one.
Thus classical physics (unlike quantum mechanics)  does $not$  require a split between  the system 
to be observed,  and the observing apparatus. And the basic ontological "entities"  of classical 
physics  are $point-particles$  and $fields$ (waves) , both moving in 4D space-time.  

\subsubsection{Determinism}

The laws of  classical physics are $deterministic$ and causal, namely the equations of motion with 
specified forces  predict  both the past and future, subject only to the initial conditions to any  
desired order of accuracy.  It is another matter that the classical behaviour of a large number of  
classical particles  is for  practical purposes more conveniently described by   "statistical"  mechanics,  
but there is no epistemological aspect  to such a strategy. A further set-back to  determinism   arises from  
$chaotic$ classical dynamics  which avers that arbitrarily close phase points can  diverge away  
exponentially from each other under dynamical evolution,  so that even carefully determined initial  
conditions do not necessarily  guarantee the future behavior to any desired accuracy. An apparent  
loss of  determinism also comes  from Brownian (random) motion in a fluid  due to collisions by 
a  large number of molecules in a fluid !    
An interesting observation ( Singh [5a]) is that Einstein's expression for the root mean square displacement  
of a Brownian particle has the same algebraic structure as Heisenberg's $UP$, except that it involves the 
diffusion coefficient of the fluid,  instead of $\hbar$, in it.  

The laws of classical physics are centred around a commonsense concept  of "locality"  whose significance is  
best manifested  by  our  capacity  to  deal meaningfully with the external world in a piecemeal fashion, and not 
all at once. Indeed,  it is enough to  identify  independent subsystems of  the  external world  to any 
desired order of accuracy,  and deal only with them,  while ignoring the rest,  since any two 
subsystems which are too far apart,  do not affect each other appreciably.   And locality  and  determinism  
in  turn  are crucial to  our concept   of "classical realism" [4] whose basic tenets are as follows : \\  
C1) All physical attributes of an individual object have definite values associated with them at any instant of time 
$irrespective$  of  their actual  measurement (which is necessarily non-invasive). \\
C2) Realism in the classical macro-world is intimately linked to $Causality$, in the sense that  the  values of 
physical attributes of an individual object at different instants of time are uniquely connected by the relevant laws.   

\subsection{Quantum  Reality: EPR Theorem [6] }

In quantum mechanics,  on the other hand,  if two systems have once interacted together, and later separated, no matter 
how far, they can no longer be assigned separate state vectors. A famous example is a spin-zero object at rest, breaking up 
spontaneously into two fragments, $A_1, A_2$ with spins  $S_1, S_2$  respectively, moving in opposite directions.  
By conservation of angular momentum,  the two spins must be equal and opposite, so that  any measurement  of one 
(say $S_1$) will automatically fix the value of $S_2 $, even without any explicit measurement !  This situation  goes much against 
intuition,  since a physical interaction between these two  objects,  receding far away from each other,  is negligible.  According to 
Schroedinger [5], this is a  most  important characteristic of quantum mechanics. ( In the modern jargon, this effect is known as 
" Entanglement "  [5b] ; see further below ). And in a famous 
paper entitled " Can quantum mechanical description of Reality be considered complete (?)", 
EPR [6, 5] considered this paradoxical aspect of quantum mechanics, by sharpening it via two  definitions: \\
i) A necessary condition for the completeness of a theory is that " every element of the physical reality 
must have a counterpart in the physical theory". \\
ii) A sufficient condition to identify an element of physical reality is " if without in any way disturbing  
the system, we can predict with certainty the value of a physical quantity, then there exists an element 
of physical reality corresponding to this physical quantity". \\
The result of these considerations was the EPR Theorem [6] , namely, the $incompatibility$  of the   
following  two statements: \\
1) the description in terms of the $\psi$-function of quantum mechanics is {\underline complete}; \\
2) the real states of spatially separated objects are independent of each other \\ 
The second statement  is called the "Einstein locality postulate".  But  then the first  statement  which says that quantum mechanics 
is a complete description, is clearly incompatible with Einstein locality. This incompatibility  is called the EPR theorem.  And,  the 
correlations in Quantum mechanics, which  are necessarily $non-local$ inasmuch as they  do $not$ decrease 
with distance, are due to {\it quantum entanglement} which violates Einstein's locality.    
\par
Thus there is a conflict between classical and quantum realisms, which calls for a more precise 
formulation pending an experimental test.  To begin with,  corresponding to  the classical realism  
tenets C1  and C2 [4]  given above, analogous  tenets Q1 and Q2 of quantum realism  may be defined as follows [4]: \\ 
Q1) Reality cannot be associated with the {\it unobserved dynamical arrtibutes} of microphysical entities; \\ 
Q2) It is not in general possible to determine the "state" of a  Quantum system, or the values of its dynamical  
attributes without affecting the system's  subsequent time evolution. Thus a measurement on a quantum 
system is necessarily $invasive$.  Later, in Sect. 5,  we give a fuller discussion on this subject in the context 
of the celebrated Bell's Theorem [7].  

\section{ Quantum  Entanglement \& Contextuality} 

The invasiveness of a quantum measurement, in turn, necessitates  some formal concepts  which 
arise in the description of  "quantum ontology" [4, 5(a) ]  in terms of  some "classical " models of quantum reality,  
proposed since  the "Copenhagen Interpretation" .    Two principal concepts associated with  the new 
paradigm   are  those of  i)  quantum  entanglement [5(a, b), 8-9] , and ii)  "quantum contextuality " [10],   as  
prerequisites  for  a description of this subject.   

\subsection{Quantum Entanglement [8 - 10]}

This is  an  aspect of quantum  theory,  which is already implied in the EPR theorem [6], 
but has come to the fore because of its  simultaneous linkage to measurement problems as well as 
to  quantum non-locality. Indeed the discussion in connection with the EPR theorem [6] above  already  
shows  that  if two systems interact, then no matter how far they get separated, their states  
get inevitably $entangled$ (non-factorizable) [9], leading to a non-trivial quantum measurement 
problem  on the  following lines  [10].  
\par
Consider a system initially in a state $(a\psi_1 + b\psi_2)$ as a superposition of two states  
$\psi_1, \psi_2$ which are the eigenstates of a dynamical variable to be measured. The interaction 
of this system with a measuring device, results in a final state of the form 
$$ \Psi = a \psi_1 \Phi_1 + b \psi_2 \Phi_2  $$ 
where $\Phi_1$ and $\Phi_2$ are two mutually orthogonal and macroscopically distinguishable 
states of the device. This final state $\Psi$ is necessarily $entangled$ [8-10],  which implies that no 
separate state can be ascribed to  the apparatus ( whose function is to register the outcome of a 
measurement by associating it with the property  of the apparatus).  Now $\Psi$ is a pure state, i.e., 
it corresponds to an ensemble of identical members,  each of whose members is described by the  
$same$ wave function $\Psi$, with a $single$ outcome of measurement.  On  the other hand, all 
measurements result in a final ensemble of systems coupled to apparatuses corresponding to 
$different$  outcomes of measurement,  implying thereby that a post-measurement ensemble 
is necessarily $heterogeneous$, i.e.,  $\Psi$  is a $mixed$ state.  
\par
Here is a genuine problem: How to generate a "mixed" state from a "pure" state  within  the   
formalism of standard quantum mechanics, since under no unitary time-evolution can a pure 
state evolve into a mixed state ! Since on the other hand,  a mixed state is a necessary condition  
for the occurence of definite outcomes of measurement, which  are $distinguishable$ from other 
outcomes,  it must be different from a "pure" state of the form $\Psi$ above.  The problem of  its `realization'   
is thus the essence of the Quantum measurement problem whose acuteness was highlighted 
by Schroedinger through his famous " Cat Paradox " [8].  In the next Section  we attempt to 
outline several approaches for circumventing  this problem, (starting from Bohr !), and arrive at  
some  workable methods  for quantum ontology.     

\subsection{  Quantum Contextuality \& Non-Contextuality}  

 "Quantum  contextuality " which   characterizes  a quantum measurement,  
 arises as follows [11] . A  quantum state vector $| \psi >$ specifies the 
probability $| <\alpha | \psi >| ^2$ of observing the set of eigenvalues $\{\alpha\}$ of a 
complete set of observables   $A$  in the experimental situation or "context"  where $A$ is observed. 
Equally, $| \psi >$ specifies the probabilities  $| <\beta | \psi >| ^2$ for observing the 
eigenvalues $\{\beta\}$  of a different set  $B$ in the " context"  where $B$  is measured. Each context 
corresponds to the experimental arrangement  to measure one complete set of observables. Due to this $inherent$ 
context dependence, quantum mechanics does $not$ specify  joint probabilities of non-commuting 
observables.  It is  usually 
assumed that $A$ and $B$ cannot be simultaneously measured if they contain mutually non-commuting 
observables,  which means that  $| <\alpha | \psi >| ^2$ and  $| <\beta | \psi >| ^2$  
refer to $different$ "contexts".  Thus quantum probabilities are inherently context -  dependent.  Further,    
the context dependence is $irreducible$, i.e., quantum mechanics cannot be imbedded  in  a classical  
"context - independent "  stochastic theory. . This is the essence of the Gleasson-Kochen-Specker theorem [12-13]. These contextuality theorems circumscribe the extent to which dynamical variables in quantum mechanics 
can be ascribed simultaneous " Reality " ,  independent of observations. 
\par 
Similarly  the  hypothesis  of  " non-contextuality"  is the assumption that  the outcome of a measurement  
of a   dynamical variable, say $A_1$,  is taken to be the same  irrespective of  any other observable   
(commuting with $A_1$) measured with it [14 ].  Such variables have been given the name " beable"  
in Bell's terminology [15],  whose values are independent of the context of observation, whereas other 
variables may have context-dependent values.  

\subsection{Quantum Locality vs. Non-Locality}  

Unlike classical locality (Sect 3.1.2), which does not extend beyond a common-sense definition, 
quantum locality  is  much harder to define  since it  is inextricably linked with non-locality effects.  
Roughly speaking,  quantum non-locality denotes quantum mechanical  action at a distance  (termed `spooky' by Einstein) 
in a situation where a distant influence appears to be counter-intuitive because of the absence 
of any classically describable form of physical mediation [10]. 
\par
More precisely, a non-local effect means affecting the state of an individual entity by any form of 
dynamical intervention in a faraway region such that no known physical influence can causally 
connect the occurrences in $that$ region to the system under study [10].  For a space-like separation, 
a quantum non-locality necessarily  implies entanglement  for spatially separated particles. 

\section{ Quantum Ontology  }

The  various  "interpretations" of  the quantum  formalism  fall broadly into two classes, I) those that are  
supplemented by extra ingredients (necessitated by the compulsions of  preparing a "mixed state" [10] !)  ;  
and II)  those that are not.  Class I includes the Copenhagen   
group of interpretations ( Bohr, Heisenberg,   von-Neumann et al), as well as the de-Broglie - Bohm  causal  
interpretations. Class II includes   Everett's relative - state and many-world approaches, as  well as the  
Quantum History approach  initiated by Griffiths, Omnes, Gellmann \& Hartle, and Zurek [5].  

\subsection{Class I Ontology : Bohr}
 
Bohr took  an  intensely pragmatic view, ( based on  logical  priority),  by giving a workable definition  
for  the $measurement$ process. He insisted that the language of classical physics is the only one  
available for communicating the observed results of any phenomenon, so that all measuring devices  
must be described by classical physics as a {\it matter of principle}.  This amounts to supplementing  
the formalism of quantum mechanics with the addition of classical physics to describe the measuring   
apparatus. Thus [5a] the combined system $S$ and measuring apparatus $A$  is an unanalyzable whole 
phenomenon $(S + A)$.  For example $(S + A_1)$  representing the electron $S$ and its position recorder   
$A_1$, is different from $(S + A_2)$ standing for the electron $S$ and its momentum recorder $A_2$,  
so that both cannot be combined into a single description corresponding to an electron with   a  known  
position and momentum. This is how the uncertainty principle is incorporated in practice: the two  
descriptions $(S+A_1)$ and $(S+A_2)$  are complementary,  The quantum world is probabilistic,  and  
a measurement  does not reveal any pre-existing property of the system.  
Bohr was against drawing any further  ontological picture  of the quantum world,   beyond the   
measurement process.  And due to the overwhelming success of the New Quantum Mechanics,   
this {\it Copenhagen Interpretation}  of Bohr had  many more followers (who were  more anxious to work out    
the multi-dimensional  consequences of the new theory ) than  those sharing Einstein's  misgivings,   
so that the  conceptual  issues emanating from the sense of incompleteness of quantum mechanics never got  
off the ground until rather recently,  when the prospects of experimentation have become brighter.

\subsection{ Class I Ontoligy : Heisenberg \& Von Neumann}

Bohr did not assign any particular significance  to the wave function $\psi$ of the quantum system in his 
" No Ontology "  view.  In contrast,  in Heisenberg's   formulation,  the wave function $\psi$ plus all observables  
$O(t)$, represents "objective tendencies" for actual events to occur. Thus while  the wave function $\psi (t)$    
represents the  wave - like aspects of nature,  the particle - like aspects are represented by " actual events ". 
This  implies the use of the concept of the "collapse of the wave function"  :  The actualization of the tendencies 
happens  when the experimental result is recorded, e.g., a click in the geiger counter. In the words of  
Heisenberg [5a], "  The observation itself changes the probability function discontinuously; it selects, of all 
possible events, the actual one that has taken place ", i.e., the wave function has "collapsed " , a 
concept that Von Neumann articulated more explicitly.  
\par
In von Neumann's  view, the measuring apparatus $A$ as well as the system $S$ are both to be described  
by quantum mechanics.  Thus one has now two kinds of eigenfunctions and eigenvalues, one for the 
system $S$ [$\psi_n $, $\omega_n$] , and one for the apparatus $A$ [ $f(a_n)$, $a_n$ ].  Von Neumann  
now postulates the " measurement interaction " which causes the initial system-apparatus state $ \psi_n f(a)$  
to evolve into $\psi_n f(a_n)$.  Then by the superposition principle of quantum mechanics, if the system is  
initially in a   superposition $\sum_n c_n \psi_n$, the measurement interaction will cause the evolution :
$$ \sum_n c_n \psi_n f(a)  \Rightarrow  \sum_n c_n \psi_n f(a_n)  $$
As a result, the probability of observing the pointer reading to be $a_n$, corresponding to the system being 
in the state $\psi_n$, will be $ |c_n|^2$. 
\par
So far, it is only quantum mechanics, giving the state $\sum_n c_n \psi_n f(a_n)$ without a specific value 
of the pointer reading. Von Neumann now postulates that when the measurement is completed, the wave 
function now $collapses$ to a $single$ term, namely, $\psi_N f(a_N)$. It is this collapse postulate that  
acts as the extra ingredient to quantum mechanics. 

\subsection{de Broglie-Bohm Causal Interpretation}

While we are still on  Class I  types of Quantum Ontology,  we digress  into  a special  type of  
the latter   for a more detailed presentation  since it  has  been the subject of  very extensive  
investigations during  recent times.  It started with  Louis de Broglie  who,  in 1927,  proposed a  
realistic causal interpretation of quantum mechanics  [16] in his  " pilot wave " theory, but it did not   
find  ready acceptance in the  smug "Bohr-filled"  atmosphere prevailing at the time. Then  in 1952,  
David Bohm [ 17 ] came up independently with a similar proposal, overcoming the earlier objections 
to the de Broglie proposal, and found support from no less a person than John Bell,  presumably because  
it  tied up with his concept of "beables" [15]. [ The  requisite "beable" in this case is the $position$ 
variable  $x$  with a special status ].  After Bell's support, there  has necessarily been   intense  
activity  in this field,  with some major Indian contributions [18-21].  We 
summarise some essential features of this theory, together with recent developments  
\par
The ontology of de Broglie - Bohm theory is "realistic" [5a].  This means that a quantum object, say 
an electron, has both a particle aspect ( a trajectory $q(t)$ associated with it), and  a wave aspect viz., 
 its wave function $\psi(q, t)$ is involved in determining its velocity ${\dot q}(t)$ 
through a relation of the form 
$$ m {\dot q}(t) = \nabla S  $$ 
where $S$ is the phase of the wave function $\psi$ and the time evolution of the latter is described 
by the Schroedinger equation. Thus the usual quantum formulation is $supplemented$ by the addition  
of particle trajectories, so as to make it "deterministic".  Which means that, given position $q(t)$ and 
wave function $\psi(q, t)$, at $t=0$, it can be predicted at all later times.  Then what is the significance 
of quantum probabilities in this scenario ? They arise through our inability to precisely control 
particle positions.   Indeed,   the best we can do is to prepare a statistical ensemble at $t=0$ in which 
the particle position is distributed according to the probability distribution 
$$ P (q, t =0 )  = | \psi (q, t=0 )|^2  $$ 
and subsequently the probability distribution evolves  in accordance with  the de Broglie-Bohm 
dynamics, namely 
$$ P ( q, t ) = | \psi (q, t ) |^2  $$ 
How does this scenario gel with the standard wave-Particle duality ?  The answer is that it is 
$not$ a particle or wave according to the experimental set-up ( a la standard quantum mechanics ), 
but that  it is  $both$ a particle and a wave. In the example of the double slit experiment, the "particle"  
of course goes through one of the slits, but the "wave" goes through both.  
\par
It should $not$ be concluded from this that we have reverted to classical  determinism, since the 
electron trajectories in the de Broglie-Bohm theory are quite different from those in standard 
classical physics.  Thus, e.g.,  in the double slit experiment, the trajectories never cross the 
middle plane between the slits !  To put it crudely though,   this theory appears as a sort of  `insertion'  
(by hand) of an element of  classical mechanics into the  quantum picture.  

\subsubsection{ Recent Developments : Roy-Singh Approach}

The above is a broad outline of the de Broglie-Bohm "causal" theory, which  also has many non-classical features, as the above discussion shows.     
In recent years it has  been generalized in several ways, but before discussing such 
generalizations, it is pertinent to recall Wigner's  famous $(q, p)$ distribution [22] based 
on standard quantum mechanics. Unlike de Broglie-Bohm theory,  the Wigner distribution gives a $symmetrical$ 
treatment to the position and momentum variables, albeit  at the cost of positive-definiteness,  as has been shown by several people [23].  Roy and Singh [18 ]  set about formulating a generalization of the de Broglie-Bohm theory designed to give a symmetrical 
treatment of position and momentum, while bravely preserving  the positivity condition on the 
joint $(q, p)$ distribution function, ( through suitable  
$\delta$-functions),  wherein the apparent violation of the above theorem [23] is sought to be reconciled via  some further insertions (by hand) on the deBroglie-Bohm picture. Roy-Singh have termed  their formulation  as  a "maximally  realistic"   description  
of  causal quantum mechanics. 

\subsection{ Class II Ontology : Everett's  " Many Worlds"} 
  
With no extra assumptions beyond quantum mechanics,  Everett  proposed an interpretation in which the Schroedinger equation provides a complete  
description of Nature [24].  Since the  measuring apparatus is described by quantum mechanics, there is no " collapse " of the wave 
function in this approach. It is a unitary description without any split between the system and  
the apparatus. B. S. de Witt [25] popularised this interpretation as the " many worlds interpretation " 
of quantum mechanics.   The idea is broadly as follows [5a] : \\
Because of the time evolution implied in the Schroedinger equation, an observer 
would be expected to be in a superposition of wave functions describing different eigenstates.  
If, e.g., we are in a normalized state 
$$  \Psi  =  c_1 \psi (walking \quad east)  +  c_2 \psi (walking \quad  west)  $$, 
as a result of the interaction, we should be simultaneously walking to the east  as well as to 
the west.  Everett asserts that this is precisely what actually happens : Due to the measurement 
interaction, we split into $two$ editions of ourselves; one walking to the east, and the other walking  
to the west !  Thus the universe has split into $two$, with probabilities $|c_1|^2$ and $|c_2|^2$ 
respectively.  But we are not aware of this split, as we seem to live in only $one$ universe. \\ 
The theory maintains objective reality, but only at the cost of a vast proliferation of the  
universes, since the universe splits every time an observation is made. On the other hand,  the wave function 
never collapses as an observation is made; only its different components inhabit different universes. 
And despite extravagant (and wasteful)  use of universes, it has an economy of principles, since 
the only input  is  the Schroedinger equation. \\
This interpretation has found an appeal to those working in Quantum Computation,  
since  [26]  the massive parallel computing associated with quantum computation fits in with the  
language of  the superposition 
$$  \Psi  = \sum c_j \psi_i   $$ 
if we regard that a computation on each $\psi_i$  is carried on in a different universe.      

\subsection{Quantum History Approach [27-30]}

This approach, which is associated with the work of Griffith [27], Omnes [28], Gell-Mann -   
Hartle [29], and Zurek [30], may be regarded as a  sort of minimal (logical) completion of the   
Copenhagen  approach,  to make up an organic whole. Its essential ingredients  may be 
summarized a la [5a]  as  follows :  \\
1) A possible set of fine-grained histories $h$; \\ 
2) A notion of coarse graining;  \\
3) A decoherence functional $D(h, h')$ defined for each pair of histories $h$ and $h'$, satisfying 
standard mathematical properties. \\
4) A Superposition principle for coarse graining defined as a standard sum over histories $h, h'$.  \\
5) A decoherence condition for two histories $h, h'$. \\ 
6) A  natural definition of  the probability $p(h)$ for the history $h$ as $ p(h) = D(h, h)$.   \\

How does the `quantum histories approach' solve the measurement problem ? It circumvents the problem of wave function collapse 
by asserting that " the measurement problem stems from the presence of those parts of the wave function corresponding to those alternatives that do not actually happen " [5b].  And when can we ignore the alternative parts of the wave function ? The answer is equally simple :  The other parts of the  
wave function can be ignored " at exactly the moment when they have no further effect on us " [5b]. From this point on, the future history of the wave 
function $decoheres$ [5b] .  Thus the interpretation provides for quantum probabilies only for histories in $decoherent$ sets. 
  
 \subsection{ Anything Else ?  Afshar Experiment  \& After}

The galaxy of " interpretations"  described above for quantum theory,   all   swear   by  the $mathematics$ of the Shroedinger Equation or its 
relativistic (Dirac) generalization, on which there is no controversy from any quarter. The `problem' seems to arise when the different 
$interpretations$, from the Copenhagen   to the Many Worlds interpretation, seem to co-exist !  Now while the Bell formulation (see next Section) 
marked a big step towards distinguishing between the `classical' vs `quantum' predictions -- and found an $experimental$ resolution of the issue, 
thanks to the Aspect experiment --,  so far no corresponding experimental way to distinguish among the  different  `intellectual'  candidates for physical interpretation of the $same$  mathematics of the quantum theory,  seems not to be forthcoming. And without such a concrete mechanism,  there seems to be  no way to prevent further proliferation of  more such alternatives.  Against this background,  the latest  Afshar experiment  [PRL 2004]  has   opened  up a fresh opportunity for a possible  resolution of the issue. 
\par
The main thrust of the Afshar experiment\footnote{as narrated in  J.G. Cramer,  http://www.npl.washington.edu/TI}, 
 is to posit a concrete test of  the  Copenhagen Interpretation  on  the impossibility of  observing both particle and wave properties in the same  experiment. This he does  through an ingenious  variant of the famous two-slit experiment  wherein he inserts one or more thin  wires at the previously measured positions of the interference minima to define  the $which-way$ routes for  the passage of the photons through a pair of pinholes, in an otherwise identical set-up to the canonical arrangement.  In one such setup,  if the wire plane is uniformly illuminated, the wires absorb about 6\% of the light. Then Afshar measures the difference in the light received at the pinhole images with and without the wires in place. What should he observe ? Both the Copenhagen and Many Worlds interpretations expect the wires to  intercept about 6\% of the light they do for uniform illumination. But actually Afshar observes almost $zero$ \% interception, for,  wires  placed at the zero-intensity locations of the interference minima intercept no light, thus implying that  the wave interference pattern is still present  despite an  unambiguous "which-way"  definition  of the passage of light photons through a pair of pinholes,  a feat hitherto considered to be impossible[ Neils Bohr, Nature {\bf 121}, 580 (1928)].  
But from this one need $not$  jump to the conclusion  that the mathematical theory of quantum mechanics has also been falsified. For, a  simple quantum mechanical calculation using the standard formalism shows that the wires should intercept only a very small fraction of the light.  So while the formal mathematics of quantum mechanics is fully vindicated, it is  these ` interpretations ' of quantum mechanics  that may need refinement, unless 
specifically tailored to meet the objections [ J. Cramers,  Reviews of Modern Physics {\bf 58}, 647 (1986)] ! 

\section {Bell's  Theorem [6]; [10]}

After this short detour on quantum ontology, we come back to a systematic  study  by John Bell 
of the consequences of the EPR theorem [6].  The result of this study which goes by the name of    
Bell's theorem  [7], is the most famous legacy of John Bell, as it draws  an important line  between quantum mechanics (QM) and the world as we know it intuitively. Despite its simplicity  and elegance,  it touches upon many of the fundamental philosophical issues that relate to modern physics. In its simplest form, Bell's theorem, which has been called  "the most profound in science" (Stapp, 1975), states: \\
No physical theory of local hidden variables can ever reproduce all of the predictions of quantum mechanics. \\
\par
To appreciate the significance of this terse Theorem it is first necessary to analyse the ingredients which Bell harnessed to arrive at this result. 
For the background,  Bell examined both the EPR  paper [6]  and  John von Neumann's  1932 proof of the incompatibility of hidden variables with QM. As to the former, since Einstein's locality postulate appeals instantly to intuition,  Bell's analysis  [7]  was 
designed  to  make the concept of Einstein locality more precise by introducing  " hidden  
variables " as a means of circumventing the counter-intuitiveness implied in  quantum entanglement ;  on the other hand, he was also 
conscious of Von Neumann's "no go theorem" on quantum mechanics  vis-a-vis hidden variables.  Taking a 
generalized view of `hidden variables', he set about formulating the ramifications of the EPR theorem in   the form of 
certain mathematical inequalities (called "Bell's inequalities") as a concrete expression of the requirement  that the assumption of local 
realism - that particle attributes have definite values independent of the act of observation, and that distant objects do not exchange information faster than the speed of light - leads to, in respect of  certain types of phenomena which do not exist  in quantum mechanics.  And one way to 
define `hidden variables' is to say that they are synonymous with  the  well defined properties associated with these inequalities. 
\par
The " inequalities"  concern measurements made by remotely located observers (often called Alice and Bob ) on entangled  pairs of particles that have interacted and then separated. Assuming hidden variables, they lead to strict limits on the possible values of the correlation of subsequent measurements that can be obtained from the particle pairs. Bell discovered that these limits are outside the predictions of quantum mechanics in special cases. Quantum mechanics(QM) does not assume the existence of hidden variables associated with individual particles, and so the inequalities do not apply to it. The QM predicted correlation is due to quantum entanglement of the pair, with the idea that their state is not determined until the point at which a measurement is made on one or the other. This idea is fully in accordance with the Heisenberg uncertainty principle , a basic tenet of  QM.
\par
If one accepts Bell's theorem: either quantum mechanics is wrong, or local realism is wrong, since they are mutually incompatible. To address the issue  scientifically, experiments needed to be performed over the  years,  with  concomitant  improvements in technology.  Bell's  test experiments  to date overwhelmingly show that the inequalities of Bell's theorem are violated [31]. This is taken by most physicists as providing empirical evidence against local realism, while constituting positive evidence in favor of QM;  but  the principle of special relativity  is  saved by the no-communication theorem , which proves that it is impossible for Alice to communicate information to Bob (or vice versa) faster than the speed of light.

\subsection{Bell's Inequalities}

The above  is as far as the  qualitative logic of Bell's Theorem goes. For a more quantitative description,   
the following sketch from Home [10] is instructive.  Consider two spin- $1/2$ particles in a singlet state,  
and moving in opposite directions  towards two measuring Stern-Gerlach magnets.  The corresponding  
wave function  is an "entangled" (non-factorable) state [10]: 
\begin{equation}
\label{1}
|\psi > = [ |\alpha >_1 |\beta >_2  -  |\beta >_1 |\alpha >_2 ] / \sqrt {2}
\end{equation}
where $\alpha$ and $\beta$ are spin-up and spin-down states respectively. Measurements of the 
components of spins ${\bf s}_1, {\bf s}_2$ along two different directions are performed on these 
particles.  For  particle 1, let ${\hat a}$ and ${\hat a}'$ be the unit directions along which  its spin 
is measured; then $A = 2 {\hat a}.{\bf s}_1$, and $A' =   2 {\hat a}'.{\bf s}_1$  are its spin components 
in these directions,  whose measured values are $\pm 1$.  Similarly for the particle 2, the quantities 
$B = 2 {\hat b}.{\bf s}_2$, and $B' =   2 {\hat b}'.{\bf s}_2$  are the corresponding spin components 
whose measured values are again $\pm 1$. 
\par
Now consider the linear combination $ AB + A'B + AB' - A'B' $.  For any given pair of spatially   
separated  particles 1 and 2 in the  singlet state, eq.(1), one can measure only $one$  of the   
products AB, A'B, AB', A'B'.  In each case  the answer must be $\pm 1$.   
The experiment consists in making measurements on a large number 
of such pairs, with the setting on one wing (particles 1) alternating between ${\hat a}$ and ${\hat a}'$; 
and on the other wing (particles 2) between ${\hat b}$ and ${\hat b}'$. In this way an ensemble of 
measurements of each of the quantities $ AB, A'B, AB', A'B' $ is performed, and the final experimental 
data are their  "average values "  $ < AB> , < A'B> ,< AB'> , < A'B'>$.  
\par
To evaluate the average values, Bell   made  two assumptions : {\bf A} an individual outcome of a 
particular measurement is definite, and not affected by what outcome is obtained by measurement  
in a region which is sufficiently separated from the entity under study. [This is the "locality" 
condition for individual events] ;  {\bf B} the randomly chosen sample of pairs on which a quantity 
like AB is measured, is typical of the entire ensemble.  These two assumptions together lead to 
a testable  constraint on the correlation functions, {\it without any input from quantum mechanics}. 
Indeed, assumption {\bf A} implies that one can associate definite measured values with both the   
spin components A($\pm 1$) and A'($\pm 1$) for each particle 1; these values are independent of  
whether or not B or B' is measured on particle 2. Similarly for B,B' vis-a-vis A, A'. Therefore each 
particle pair has a value $\pm 1$ of each of the measured quantities  $ AB, A'B, AB', A'B' $.  Then, 
for each of the 16 different cases corresponding to the possible choices $\pm 1$ for each of 
$A, A', B, B' $ separately, the  resultant of the pair combinations gives 
\begin{equation}
\label{2}
AB + A'B + AB' - A'B'  =  \pm 2
\end{equation}
where it is understood that both the occurrences of, say, $A$, in the lhs of (2), have the $same$ value; 
and similarly for $A', B, B' $. [This is the locality condition once again]. Summing (2) over the entire 
ensemble of pairs, and taking the average, one gets
\begin{equation}
\label{3}
| < AB> + < A'B> + < AB'> - < A'B'> |   \leq 2
\end{equation}
where, by virtue of {\bf B}, the various averages can be identified with the experimentally measured  
values of the corresponding correlated quantities. This is Bell's inequality which is a verifiaable 
prediction involving measurable quantities, as a direct consequence of the locality condition.

\subsection{Quantum Violation of Locality}

This inequality is violated by  the quantum mechanical results for the singlet state which yields 
$<AB>$ = $-{\hat a}.{\hat b} $ = - $ cos \theta $.  The maximum violation occurs when all directions 
are coplanar,  with ${\hat a}'.{\hat b}'$ = $ - 1 / \sqrt {2}$, and the other 3 cosines are  all $1 /\sqrt{2}$. 
For these values the rhs equals $ 2 \sqrt{2}$.  There are of course a whole range of angles over 
which the above inequality is violated quantum mechanically. 
\par
What is the significance of this result ? It is that, irrespective of the specifics of any causal model, 
no causal theory satisfying the locality condition,  can be fully consistent with the formalism of 
quantum  mechanics -- a  classic  "no-go theorem" !    Indeed  a decisive experimental refutation [31]  of Bell's inequality  has a non-trivial  
implication of {\it irreducible quantum inseparability for macroscopic separations}. This is a 
repudiation of the cherished notion of  the principle that  any composite extended system may be 
regarded as composed of elements which are localized in separate regions,  in favour of an 
indivisible whole ! 

\section{Responses to Quantum Violations}

Possible responses to  quantum violations of Bell's theorem are: a) give up the assumption that 
$\psi$  provides a complete description of the state of an individual entity; b) a viable causal model  
of  quantum mechanics  must be $non-local$;  c)  Other forms of locality.  Possibility (a) 
points to  the need for  modifying  the formalism, say  with  "hidden" variables to further specify  the  
state of an individual microsystem, so as to make it compatible with the quantum measurement  
problem.  Possibility (b) raises the question of compatibility with Lorentz  invariance.  

\subsection{Hidden Variable Effects}

As  response (a) to the quantum violations of Bell's theorem, we may  incorporate  hidden variable  
effects, within the above formalism [10].  Denoting  hidden variables  by $\lambda$, 
the amplitude $A$, e.g., may be represented by   $A({\hat a}, \lambda)$,  to show its 
dependence on $\lambda$;  similarly for the other amplitudes $A', B, B' $.  Then  the   
result of measurement $A({\hat a}, \lambda)$  of the ${\hat a}$ spin-component of the first particle 
${\bf  2s}_1.  {\hat a}$ would be $ \pm 1$, i.e.  $A({\hat a}, \lambda)$ = $\pm 1$.  Similarly, the   
result of measurement $B({\hat b}, \lambda)$  of the ${\hat b}$ component of spin of the second  
particle ${\bf  2s}_2 . {\hat b}$ would be  given by $B({\hat b}, \lambda)$ = $\pm 1$.  Now Einstein  
locality implies that $A$ does not depend  on ${\hat b}$, nor does $B$ depend on ${\hat a}$.   
One further  stipulates  that  the measurement epochs  are such that no direct light signals   
can travel between the two Stern-Gerlach magnets. 
\par
The spin correlations are now given by the average value $<{\hat a}, {\hat b}>$ of the product $AB$ :
\begin{equation}\label{4} 
<{\hat a}, {\hat b}>  = \int d\lambda \rho(\lambda)  A({\hat a}, \lambda) B({\hat b}, \lambda)  
\end{equation} 
where $\rho(\lambda)$ is the non-negative normalized probability distribution of the hidden variables 
for the  given quantum mechanical singlet state $\psi$;  and [5a]
$$ | A({\hat a}, \lambda) |   \leq 1 ;   | B({\hat b}, \lambda) |  \leq 1  $$ 
From these representations  one shows the Bell's inequalities [5a] 
$$ | <{\hat a}, {\hat b}>  -  <{\hat a}, {\hat b}'> |  +  | <{\hat a}', {\hat b}'> + <{\hat a}', {\hat b}> |  \leq 2 $$  
On the other hand, quantum mechanics predicts 
$$   <{\hat a}, {\hat b}> = <{\bf  2s}_1. {\hat a}  {\bf  2s}_2.  {\hat b}> = - {\hat a}. {\hat b} $$ 
This last form of  $<{\hat a}, {\hat b}>$, whose modulus usually exceeds $2$,  cannot  therefore  
satisfy the Bell's inequality for a arbitrary choice  of directions. Then it follows that no local hidden   
variable theory can reproduce all the results of   quantum mechanics. This is once again    
Bell's Theorem.    
\par
Bell's inequalities are tested by "coincidence counts" from a Bell test experiment.  Pairs of particles are emitted as a result of a quantum process, analysed with respect to some key property such as polarisation direction, then detected. The settings (orientations) of the analysers are selected by the experimenter.
The source S produces pairs of "photons", sent in opposite directions. Each photon encounters a two-channel polariser whose orientation (a or b) can be set by the experimenter. Emerging signals from each channel are detected and coincidences of four types (++, - -, +- and -+) counted by the coincidence monitor. Bell test experiments to date overwhelmingly suggest that Bell's inequality is violated [31] 
The phenomenon of quantum entanglement that is implied by violation of Bell's inequality is just one element of quantum physics which cannot be represented by any classical picture of physics; other non-classical elements are complementarity.

\subsubsection{Single Particle Correlations}

But  experiments  with  two correlated photons  need not be the  only possible way  to show quantum 
mechanical violations of  orthodox paradigms.  An interesting experiment 
was proposed by Home et al  [32] with single photon states incident on two prisms facing each other,  
designed to show   simultaneous particle and wave characteristics in apparent violation of  Bohr's  
complementary  principle in its usual form.  The performed experiment [33] verified their quantum  
optical prediction [32] .  
\par 
Coming now to Bell's inequalities,  experiments beyond two-photon correlations [31]  may also include 
material particles.  A recent   $single$  neutron interferometry experiment [34] showed violation of  
Bell's inequalities  by  measuring  correlations between $two$  degrees of freedom 
(comprising spatial and spin components) of $single$  neutrons, (thus obviating the need for a 
source of entangled neutron pairs). 
 
\subsection{ "Unnikrishnan Locality"}

A radical departure from these "orthodox"  responses to quantum violations of  Einstein locality  
is provided by a different line of thought proposed by Unnikrishnan [35] that local amplitudes with  
random initial phases can be assigned to individual particles of a correlated entangled system   
and that the quantum correlations are encoded at source in the relative phase of these amplitudes. 
This has the dramatic effect that Einstein locality is not violated during observations on entangled system,  
contrary to the widespread standard belief ! This  " Unnikrishnan locality" ( see  B. d' Espagnet [36])  
 claims to preserve Einstein locality through  its insistence  on   
{\it conservation laws}  to $predetermine$  the correlations.

\section{ Quantum Zeno Effect And All That }

We now come to a topic which  is closely related  to the measurement syndrome, but whose philosophical  implications 
can be traced  all the way to the Greek philosopher Zeno,  as it goes by the name of  the   quantum Zeno effect [37-40].   
It is a seemingly paradoxical result  in quantum theory  concerning the  slowing down  of the evolution of  a dynamical system under  
repeated observation of an unstable system over a period of time, so that its decay can be greatly inhibited.  Although 
the idea in the quantum context dates back to Schroedinger [37],   
the present  interest in the subject owes its origin  to the seminal work of Misra and Sudarshan [40].  More recently, it was predicted that repeated measurements can  even 
$enhance$ the decay [41 - 43], a phenomenon which was termed $Anti-Zeno$ effect.  The experimental observation of these effects relies on the ability 
to reset the evolution of the system during the non-exponential time of the decay .  We first  outline the Misra-Sudarshan [40] theory 
for non-exponential decays giving rise to the Zeno effect, after which we briefly indicate the possibility of  anti-Zeno effects, together with 
experimental results [44].  

\subsection{ Unstable Quantum System : Definitions [40] }

There are 3 main ingredients for the quantum description of an unstable system : i) a Hilbert space ${\cal H}$ of state 
vectors, including the unstable states and their decay products; ii) a  unitary group $U_t = \exp {-iHt}$ acting on ${\cal H}$
to describe its time evolution; iii) the subspace ${\cal M}$ of ${\cal H}$ formed by the undecayed  unstable states of the  
system.  The orthogonal projection onto ${\cal M}$ is denoted by ${\cal E}$,  a two-valued observable that corresponds to  
the "yes-no" experiments   to determine if the system is in an undecayed or decayed state respectively. In terms of these 
quantities, the probability $P(t)$  for finding the system undecayed at time $t$,  if it was prepared in the state $\rho$ at $t = 0$, 
is given  by $P(t)$ = $ Tr[\rho U_t ^* {\cal E} U_t]$ , while the probability that at the instant $t$ the system is found decayed,  
is the complementary  quantity $Q(t)$ = $  Tr[\rho U_t ^* {\cal E}^\perp U_t]$,  such that $P(t) + Q(t) = 1$, and 
${\cal E} + {\cal E}^\perp =1$. 
\par
While the quantities $P(t)$ and $Q(t)$  which correspond to  specific instants of time, are unambiguously given by the standard rules  
of quantum mechanics,  they  are by themselves  inadequate  for constructing   $P(\Delta, \rho)$  and  $Q(\Delta, \rho)$ which represent the probabilities 
that  the system prepared initially in the undecayed state $\rho$ will be found to   decay  or remain undecayed (respectively), 
{\it sometime during or throughout (respectively) a  given interval $\Delta = (0,t)$}  [45].  And till the Misra-Sudarshan paper [40], quantum theory did not 
have any ready formula for  $P(\Delta, \rho)$  or  $Q(\Delta, \rho)$, on the basis of which they were led to conjecture that 
quantum theory might well be incomplete !  They proceeded to investigate the problem as follows [40]. 

\subsection{Repeated and Continuous Observations}

They looked for the operational meaning of such probabilities over $extended$  time intervals  in terms of the outcomes of  
continuous  monitoring of the unstable particle for its existence in the undecayed states.  Now continuous monitoring may be  
regarded as a limit of frequently repeated observations as the dead time between successive observations tends to zero. 
They further argued that that there is no bar in quantum theory  {\it per se} against such idealization since if were so, it would 
imply "discreteness of time" !  They then  attempted to construct the probability $P(\Delta, \rho)$ as the limit of the 
corresponding probability under successive measurements, as the interval between them approaches zero. 
\par
As a beginning, consider only three measurements at times $t=0, t/2, t$, and seek the probability that the system initially 
prepared in a given state $\rho$ will be found  in a decayed state in at least one of these three measurements. According to 
the normal rules, this probability is the $sum$ of the following three: 1) the probability for a decayed state at $t=0$; 2) the 
conditional probability for a decayed state at $t= t/2$, given that the system was undecayed at $t=0$; 3) the conditional 
probability for decay at $t=t$, given that the two preceding measurements had left the system undecayed.  Now the  first 
probability is just $P(0)$ = $Tr [\rho {\cal E }^\perp]$. But the other two (conditional) probabilities require the knowledge of 
the {\it state changes} caused by measurements at $t=0, t/2$, involving $collapse$ of the state vector  (or reduction 
of wave packet),  caused by  the very process of measurement [ see sect. 4.2 ],  which needs a short digression. 

\subsection{Collapse of State Vector }

 The initial state $\rho$ collapses due to measurement at $t=0$, to the state $\rho' = {\cal E} \rho {\cal E}$. 
Thereafter, its evolution until the second measurement at $t=t/2$ is governed by $U_t$. Thus the system,  
after being found to be undecayed at $t=0$ is in the state $\rho''$ = $U_{t/2}\rho' U_{t/2}^*$. Hence  the  
conditional probability for finding the system in a decayed state at $t=t/2$ when it was undecayed at $t=0$  is [40]
\begin{equation}
P(t/2; 0) = Tr [ {\cal E}^\perp \rho'' {\cal E}^\perp]
\end{equation}
In the same way, the conditional probability $P(t; t/2)$ works out as 
\begin{equation}
P(t; t/2) = Tr [ {\cal E}^\perp U_{t/2}^2 \rho (U_{t/2}^*)^2 {\cal E}^\perp]
\end{equation}
The total probability for finding the system decayed is the sum of these 3 conditional probabilities which 
simplifies  after algebraic rearrangements,  to 
\begin{equation}
P(\Delta, \rho)_2 = 1 - Tr [ \rho ({\cal E} U_{t/2}{\cal E})^{*2} ({\cal E}U_{t/2}{\cal E})^2 ]
\end{equation}   
The law is now fairly clear. Therefore, generalizing the sequence, one can compute the probability for 
finding the system in a decayed state in at least one of the sequence of $(n + 1)$ measurements 
undertaken at $t = (0, t/n, 2t/n, ..., t)$ in the form 
\begin{equation}
P (\Delta; \rho)_n = 1 -   Tr [ \rho ({\cal E} U_{t/n}{\cal E})^{*n} ({\cal E}U_{t/n}{\cal E})^n ]
\end{equation}
whose limit at $n \rightarrow \infty$ is expected to be the desired probability $P(\Delta; \rho)$. 

\subsection{ Quantum Zeno Paradox }

To obtain the $n \rightarrow \infty$ limit of the above, it is necessary to study the quantity
$$ \lim_{n = \infty}  ({\cal E}U_{t/n}{\cal E})^n  \equiv T(t)  $$
Then it was shown by Misra-Sudarshan [42] that under some general conditions like semi-boundedness 
of the Hamiltonian,  
\begin{equation}
T^* (t) T(t) = {\cal E} 
\end{equation}
Substitution of this result in (9) shows that in $n = \infty$ limit, $P(\Delta; \rho)$ = $ 1 - Tr(\rho {\cal E})$,  
so that it is $independent$ of $t$ ! So if the initial state is undecayed, i.e., $Tr(\rho{\cal E}) = 1$, then  
$P(\Delta; \rho) = 0$ for all intervals $\Delta$.  Thus one arrives at the paradoxical conclusion that 
the probability for an unstable particle will be found in a decayed state at $some$ time during $(0, t)$, 
is $zero$, no matter how large $t$ is. This  is the  quantum Zeno paradox   [42],  which is more 
picturesquely expressed by the statement that an unstable  quantum system under continuous 
observation, {\it does not decay} ! 
\par
The startling nature of this result naturally raises doubts on its validity. Therefore unless a more 
reliable way than the above [40] is found for the derivation of $P(\Delta; \rho)$, the very issue of 
completeness of quantum mechanics [1] would remain open to question. Note also that the 
above derivation has made essential use of the concept of "collapse of the wave function" 
[type I ontology, Sect. 4] as part of the measurement process.    

\subsection{Deviations from Exponential Decay}

Deviations from exponential decay in quantum mechanics were predicted by Khalfin [45]  
who showed that if the Hamiltonian has a spectrum bounded from below, then the 
survival probability $P$ is $not$ a pure exponential, but rather of the form $\exp{(-c t^q)}$ 
where $q < 1$  and $ c > 0$.  More generally, Winter [46] showed that the survival probability 
begins with a non-exponential oscillatory behaviour, after which the system evolves  
according to the exponential law, and finally tapers off like an inverse power of time.  A 
similar result was found by Chiu et al [47] who showed that  the time parameters $ T_1, T_2$   
separating the above three domains bear the inequalities 
$ T_1 << \Gamma ^{-1} << T_2$, where $\Gamma ^{-1}$ is the total life-time for decay.   And  
the quantum Zeno of [40] refers to the time period $T_1$ only.  The initial non-exponential 
decay behaviour is related to the fact that the coupling between the decaying system and  
the reservoir  is reversible for short enough times, for which the decayed and undecayed  
are not yet resolvable [44].  Indeed the Misra-Sudarshan theorem [40] can also be 
understood from the fact that, given a finite value of the mean energy of the decaying state,
the survival probability $P(t)$ obeys the relation [48] 
\begin{equation}
\{d P(t) / dt \}_{t = 0}  =0 
\end{equation}  
which is a general property independent of the details of the interaction. 

\subsection{Quantum Zeno vs Anti-Zeno}

The Misra-Sudarshan prediction [40] of the quantum Zeno effect was restudied [41-43] 
by focussing on the frequency of  observations, and on the decay of an unstable system 
as a consequence of  the existence of a reservoir of possible states.  The result was the 
prediction of the $opposite$ effect .  Namely, repeated observations must shorten the 
life-time of an unstable system -- the anti-Zeno effect !   Both types were observed by 
Medina et al [44] in an unstable system in the forms  of inhibition and enhancement 
respectively, by frequent measurement during the non-exponential time.

\section{Quantum Optics \& Information [49]}

This narrative will remain incomplete without  some comments on the Quantum Information Theory that has developed dramatically over the past two decades, driven by the prospects of quantum-enhanced communication and computation systems. It exploits the unique properties of quantum mechanics which facilitate different ways of communication, processing and information.  For one thing, a quantum system must evolve coherently, which means that it must remain isolated from the external environment. But to process the (quantum)  information,  it must have strong interaction with classical measuring systems and control elements. An important property of a quantum information  is that it cannot be cloned, since any attempt to `touch' it will destroy it, 
via the Uncertainty Principle. Thus quantum entanglement is  an essential ingredient of any communication task. On the positive side,  a computer algorithm can be more efficiently processed than a classical system. Indeed, soon after Feynman's prediction (1987) of this feature,  Shor gave an algorithm for the determination of prime factors which can be processed much faster by a quantum computer than with a classical one.  A related algorithm due to Grover [50] facilitates a faster searching ( $~ \sqrt{N}$) of a data-base of $N$ items than one ($~ N /2$) by classical means.  
\par
The unit of quantum information, like the classical, is based on a binary system of quantum states (like $|0>, |1>$), and is measured in $qubits$ (information digitally encoded on a quantum system), in analogy to classical $bits$.  But there exist more possibilities in a quantum system, 
such as the bases can be $|\pm>$ which are orthogonal  combinations of   $|0>, |1>$. Further, qubits can span different bit values  simultaneously. 
This is trivially true of a single qubit, but also holds for multi-qubit states.  However,  in a quantum system, different bases do $not$ commute. 
Thus  a  measurement in one basis disturbs the bit values in another basis.  This feature can be employed for a secure communication channel via 	{\it quantum key distribution} or QKD. [This is  also called quantum cryptography]. 	

\subsection{Quantum Teleportation}

Let us illustrate the new ideas in quantum communication through the working of Quantum Teleportation, or entanglement-assisted teleportation. It  is a technique that transfers a quantum state to an arbitrarily distant location using a distributed entangled state  and the transmission of some classical information.  The key theoretical ingredients involved are EPR  driven quantum entanglement, and (suitable manipulation of) Bell states (see Sect 7).  Quantum teleportation does not transport energy or matter, nor does it allow communication of information at superluminal speed.
\par
In the standard nomenclature in quantum information: the two parties are Alice (A) and Bob (B), and a qubit is in general a  unit vector in two-dimensional Hilbert space.  Suppose Alice has a qubit in some arbitrary quantum state 
\begin{equation}
| \psi_o> = \alpha |0>_o + \beta |1>o 
\end{equation}
. Assume that this quantum state is not known to Alice,  and she would like to send this state to Bob. Simple options like i)  physically transporting  the qubit to Bob,  or.ii) broadcasting the information,   are not viable,  since fragileness of quantum states rules out (i) ;  and  a "no-cloning theorem" forbids (ii).  Again $classical$ transportation (result of her measurement communicated to Bob, who then prepares the qubit in his possession accordingly) is impossible since quantum information cannot be measured reliably.    
Thus, Alice has a seemingly impossible problem on hand.  A solution was discovered by Bennet et al. [50]:   The parts of a maximally entangled two-qubit state are distributed to Alice and Bob. The protocol then involves Alice and Bob interacting locally with the qubit(s) in their possession and Alice sending two classical bits to Bob. In the end, the qubit in Bob's possession will be in the desired state. The logic goes as follows. 

Suppose Alice has a qubit that she wants to teleport to Bob. 
Our quantum teleportation scheme requires Alice and Bob to share a maximally entangled (EPR driven)  state beforehand, for instance two  2-particle Bell states  
\begin{equation}
 |\Phi ^{\pm} >  = [|0>_A \otimes |>_B  \pm  |1>_A \otimes |1>_B ] / \sqrt{2} 
\end{equation}   

Alice takes one of the particles in the pair, and Bob keeps other one. The subscripts A and B in the entangled state refer Alice's or Bob's particle.
So, Alice has two particles (O, the one she wants to teleport, and A,  one of the entangled pair), and Bob has one particle, B. In the total system, the state of these three particles is given by: 
$ | \psi_o> \otimes | \Phi ^+ > $, with  the above definitions.   
Alice will then make a partial measurement in the Bell basis on the two qubits in her possession. To make the result of her measurement clear, we will rewrite the two qubits of Alice in the Bell basis via the following general identities expressed in pairs in a compact notation :
\begin{eqnarray} 
|0> \otimes |0> ; |1> \otimes |1> & = & [\Phi ^+  \pm \Phi ^- ] / \sqrt{2} ;  \nonumber  \\
|0> \otimes |1>; |1> \otimes |0>  & = & [\Psi^+ \pm \Psi^- ] / \sqrt{2}  
\end{eqnarray} 
The three-particle state $ | \psi_o> \otimes | \Phi ^+ > $, when expanded,  then  becomes: 
\begin{eqnarray*}
| \Phi^+>_A \otimes [\alpha |0> +\beta |1>]_B & + &  | \Phi^->_A \otimes [\alpha |0> - \beta |1>]_B  \\
|\Psi^->_A \otimes [\alpha |1> +\beta |0>]_B  & + & |\Psi^- >_A  \otimes [\alpha |1> - \beta |0>]_B  
\end{eqnarray*}
Notice all we have done so far is a change of basis on Alice's part of the system. No operation has been performed and the three particles are still in the same state. The actual teleportation starts when Alice measures her two qubits in the Bell basis. Given the above expression, evidently the results of her (local) measurement on the total system is that the three particle state would collapse to one of the following four states (with equal probability of obtaining each):
$$ |\Phi^ \pm >_A \otimes [\alpha |0> \pm  \beta |1>]_B ;  \quad  |\Psi^ \pm>_A \otimes [\alpha |1>  \pm \beta |0>]_B  $$

Alice's two particles are now entangled to each other, in one of the four Bell states. The entanglement orginally shared between Alice's and Bob's is now broken. Bob's particle takes on one of the four superposition states shown above. Note how Bob's qubit is now in a state that resembles the state to be teleported (the four possible states for Bob's qubit are unitary images of the state to be teleported).
The crucial step, the local measurement done by Alice on the Bell basis, is done. It is clear how to proceed further. Alice now has complete knowledge of the state of the three particles; the result of her Bell measurement tells her which of the four states the system is in. She simply has to send her results to Bob through a classical channel. Two classical bits can communicate which of the four results she obtained.
After Bob receives the message from Alice, he will know which of the four states his particle is in. Using this information, he performs a unitary operation on his particle to transform it to the state  $[\alpha |0> +\beta |1>]$.  
If Alice indicates her result is $ | \Phi^+>$ , Bob knows his qubit is already in the desired state and does nothing. This amounts to the trivial unitary operation, the identity operator. 
If the message indicates $ | \Phi^->$, Bob would send his qubit through the unitary gate given by the Pauli matrix  $\sigma_3$  to recover the state.
If Alice's message correspond to $| \Psi ^ +>$ , Bob applies the "gate" (unitary transformation, that is) via Pauli matrix  $\sigma_1$ to his qubit.
Finally, for the remaining case, the appropriate "gate"  is given by $ \sigma_3 \sigma_1 = i \sigma_2$. Teleportation is therefore achieved.
Experimentally, the projective measurement done by Alice may be achieved via a series of laser pulses directed at the two particles.
\par
After this operation, Bob's qubit will take on the state $[\alpha |0> +\beta |1>]$ , and Alice's qubit becomes an (undefined) part of an entangled state. Thus, teleportation does not result in the copying of qubits, and hence is consistent with the no- cloning theorem. 
There is no transfer of matter or energy involved. Alice's particle has not been physically moved to Bob; only its state has been transferred. The term "teleportation", coined by Bennett et al [51],  reflects the indistinguishability of quantum mechanical particles. 
The teleportation scheme combines two "impossible" procedures. If we remove the shared entangled state from Alice and Bob, the scheme becomes classical teleportation, which is impossible as mentioned before. On the other hand, if the classical channel is removed, then it becomes an attempt to achieve superluminal communication, again impossible, via   "no communication"  theorem.  

\subsection{ Future trends in Quantum Communication}

The foregoing merely offers a `taste'  of the nature of quantum teleportation which is a rapidly growing field. 
 Now, present-day quantum key distribution systems can operate only over the distances of several tens of kilometers, which may severely limit their practical applicability. due to various  losses and noise in the communication lines (telecom optical fibers) and detectors.  One of the major challenges in the field presently is to develop long-distance quantum communication networks which allow secure quantum communication over arbitrarily long distances. Its main ingredients -- the quantum memory for light, the source of entangled two-mode squeezed states, quantum teleportation, and entanglement swapping-- have now been experimentally demonstrated in the laboratories. So a major  task is the integration of these basic building blocks into a quantum network.  This truly interdisciplinary effort requires a close collaboration of experimentalists with theoreticians and of quantum opticians with atomic physicists. A very promising future direction is the so-called "hybrid"quantum information processing which combines the approaches developed in the fields of single-photon linear optics QIP,  and  a corresponding QIP  based on  quantum continuous variables.  The latter
seem to be particularly suitable for quantum communication since they offer large bandwidth and may be easier to manipulate than quantum bits. And,  unlike discrete atomic spins, distant atomic continuous-variable systems can be entangled and that a light-atoms quantum state exchange can be performed at the level of continuous variables. The resulting atomic quantum memory for light is thought to be a crucial component of the future quantum communication networks (see also ref [52]). 

\subsection{ Emergence of Quantum Biology} 
For sometime scientists have been toying with ideas involving quantum phenomena [53] towards understanding the logic of " Nature " that always seems to have got around to do things in the most efficient and economical of ways. Also, thanks to nanotechnology, an increased understanding of mechanisms operating at the " nano-level " seem to have emboldened a rethink of existing paradigms : to consider the possibility of biological systems utilizing " quantum wierdness " ! In a highly suggestive paper, Patel [54] has drawn a parallel to Grover's algorithm mentioned above [50] with that employed in DNA replication and protein synthesis. Of course quantum algorithmic mechanisms would require a sufficiently protected microenvironment for keeping decoherence effects at bay in the highly interactive cellular environment. Extensive experimental tests in concert with progress in our understanding of quantum information theory could someday bring out life's secrets.

\section{Retrospect And Conclusion }

The foregoing is a rather sketchy description of the state of the art   in  the emerging  field of foundations of quantum theory,  
 The central issue is the   "conceptual  anatomy"  of quantum mechanics  
(bearing on quantum non-locality  and  the measurement process) that   has  sprung up  to life,  during the last  
3 decades,  in the context of  fresh experimental  prospects for  its resolution.   
\par
Most of these efforts have been   designed to throw light on the quantum measurement  paradox, namely,  $how$ a 
definite outcome occurs  in an individual measurement, (although standard quantum mechanics predicts a coherent  
superposition of  different outcomes). A whole gamut of proposals since the Copenhagen Interpretation (see  Sect 4) have come up  
to resolve the issue. 
\par
Another pathology  is the so-called  Zeno's paradox [40] bearing on the effect of frequent measurements on the 
time of decay of an unstable system.  Since it makes essential use of the "collapse of state vector" ,  (one of 
the main contenders deemed responsible for the above dichotomy) an experimental resolution of Zeno's paradox  [44] 
may  well  hold the key to a resolution of the measurement paradox.  Finally the intriguing  concept of quantum teleportation has opened up 
a most exciting field with vast possibilities for the future. 

\section*{Acknowledgements}

I am grateful to Professors V. Singh, S.M. Roy, D. Home, B. Misra, and C. S. Unnikrishnan, for making 
 available their own writings plus allied literature,  without which the present 
document would not have been possible. I am no less grateful to Gargi M Delmotte for a perspective  on the evolving subject of quantum biology.

\end{document}